\documentclass[a4paper,12pt]{article}

\usepackage{latexsym}

\newcommand{\nc}{\newcommand}    

\nc{\be}[1]{\begin{equation}\mbox{$\label{#1}$}}
\nc{\bea}[1]{\begin{eqnarray} \mbox{$\label{#1}$}}
\nc{\Section}[2]{\section{#2}\label{#1}}
\nc{\Bibitem}[1]{\bibitem{#1}}
\nc{\Label}[1]{\label{#1}}

\nc{\eea}{\end{eqnarray}}
\nc{\ee}{\end{equation}}

\nc{\bdm}{\begin{displaymath}}
\nc{\edm}{\end{displaymath}}
\nc{\dpsty}{\displaystyle}
\nc{\bc}{\begin{center}}
\nc{\ec}{\end{center}}
\nc{\ba}{\begin{array}}
\nc{\ea}{\end{array}}
\nc{\bab}{\begin{abstract}}
\nc{\eab}{\end{abstract}}
\nc{\btab}{\begin{tabular}}
\nc{\etab}{\end{tabular}}
\nc{\bit}{\begin{itemize}}
\nc{\eit}{\end{itemize}}
\nc{\ben}{\begin{enumerate}}
\nc{\een}{\end{enumerate}}
\nc{\bfig}{\begin{figure}}
\nc{\efig}{\end{figure}}

\nc{\arreq}{&\!=\!&}
\nc{\arrmi}{&\!-\!&}
\nc{\arrpl}{&\!+\!&}
\nc{\arrap}{&\!\!\!\approx\!\!\!&}
\nc{\non}{\nonumber}
\nc{\align}{\!\!\!\!\!\!\!\!&&}

\def\lsim{\; \raise0.3ex\hbox{$<$\kern-0.75em
      \raise-1.1ex\hbox{$\sim$}}\; }
\def\gsim{\; \raise0.3ex\hbox{$>$\kern-0.75em
      \raise-1.1ex\hbox{$\sim$}}\; }

\nc{\DOT}{\hspace{-0.08in}{\bf .}\hspace{0.1in}}
\nc{\Laada}{\hbox {$\sqcap$ \kern -1em $\sqcup$}}
\nc\loota{{\scriptstyle\sqcap\kern-0.55em\hbox{$\scriptstyle\sqcup$}}}
\nc\Loota{{\sqcap\kern-0.65em\hbox{$\sqcup$}}}
\nc\laada{\Loota}
\nc{\qed}{\hskip 3em \hbox{\BOX} \vskip 2ex}

\nc{\real}{{\rm I \! R}}
\nc{\Z}{{\sf Z \!\!\! Z}}
\nc{\complex}{{\rm C\!\!\! {\sf I}\,\,}}
\def\bigid{\leavevmode\hbox{\small1\kern-3.8pt\normalsize1}}
\def\id{\leavevmode\hbox{\small1\kern-3.3pt\normalsize1}}
\nc{\slask}{\!\!\!/}
\nc{\bis}{{\prime\prime}}
\nc{\pa}{\partial}
\nc{\na}{\nabla}
\nc{\ra}{\rangle}
\nc{\la}{\langle}
\nc{\goto}{\rightarrow}
\nc{\swap}{\leftrightarrow}

\nc{\EE}[1]{ \mbox{$\cdot10^{#1}$} }
\nc{\abs}[1]{\left|#1\right|}
\nc{\at}[2]{\left.#1\right|_{#2}}
\nc{\norm}[1]{\|#1\|}
\nc{\abscut}[2]{\Abs{#1}_{\scriptscriptstyle#2}}
\nc{\vek}[1]{{\rm\bf #1}}
\nc{\integral}[2]{\int\limits_{#1}^{#2}}
\nc{\inv}[1]{\frac{1}{#1}}
\nc{\dd}[2]{{{\partial #1}\over{\partial #2}}}
\nc{\ddd}[2]{{{{\partial}^2 #1}\over{\partial {#2}^2}}}
\nc{\dddd}[3]{{{{\partial}^2 #1}\over
    {\partial #2 \partial #3}}}
\nc{\dder}[2]{{{d #1}\over{d #2}}}
\nc{\ddder}[2]{{{d^2 #1}\over{d {#2}^2}}}
\nc{\dddder}[3]{{d^2 #1}\over
    {d #2 d #3}}
\nc{\dx}[1]{d\,^{#1}x}
\nc{\dy}[1]{d\,^{#1}y}
\nc{\dz}[1]{d\,^{#1}z}
\nc{\dl}[1]{\frac{d\,^{#1}l}{(2\pi)^{#1}}}
\nc{\dk}[1]{\frac{d\,^{#1}k}{(2\pi)^{#1}}}
\nc{\dq}[1]{\frac{d\,^{#1}q}{(2\pi)^{#1}}}

\nc{\bfT}{{\bf T }}
\def\GeV{{\rm\ GeV}}

\nc{\cA}{{\cal A}}
\nc{\cB}{{\cal B}}
\nc{\cD}{{\cal D}}
\nc{\cE}{{\cal E}}
\nc{\cG}{{\cal G}}
\nc{\cH}{{\cal H}}
\nc{\cL}{{\cal L}}
\nc{\cO}{{\cal O}}
\nc{\cT}{{\cal T}}
\nc{\cN}{{\cal N}}
\nc{\rvac}[1]{|{\cal O}#1\rangle}
\nc{\lvac}[1]{\langle{\cal O}#1|}
\nc{\rvacb}[1]{|{\cal O}_\beta #1\rangle}
\nc{\lvacb}[1]{\langle{\cal O}_\beta #1 |}
\nc{\bb}{\bar{\beta}}
\nc{\bt}{\tilde{\beta}}
\nc{\ctH}{\tilde{\cal H}}
\nc{\chH}{\hat{\cal H}}
\nc{\1}{\aa}
\nc{\2}{\"{a}}
\nc{\3}{\"{o}}
\nc{\4}{\AA}
\nc{\5}{\"{A}}
\nc{\6}{\"{O}}
\nc{\al}{\alpha}
\nc{\g}{\gamma}
\nc{\Del}{\Delta}
\nc{\e}{\textrm{e}}
\nc{\eps}{\epsilon}
\nc{\lam}{\lambda}
\nc{\Om}{\Omega}
\nc{\ve}{\varepsilon}
\nc{\mn}{{\mu\nu}}
\nc{\vp}{\varphi}

\nc{\advp}[3]{{\it  Adv.\ in\ Phys.\ }{{\bf #1} {(#2)} {#3}}}
\nc{\annp}[3]{{\it  Ann.\ Phys.\ (N.Y.)\ }{{\bf #1} {(#2)} {#3}}}
\nc{\apl}[3]{{\it  Appl. Phys. Lett. }{{\bf #1} {(#2)} {#3}}}
\nc{\apj}[3]{{\it  Ap.\ J.\ }{{\bf #1} {(#2)} {#3}}}
\nc{\apjl}[3]{{\it  Ap.\ J.\ Lett.\ }{{\bf #1} {(#2)} {#3}}}
\nc{\app}[3]{{\it Astropart.\ Phys.\ }{{\bf #1} {(#2)} {#3}}}
\nc{\cmp}[3]{{\it  Comm.\ Math.\ Phys.\ }{{ \bf #1} {(#2)} {#3}}}
\nc{\cqg}[3]{{\it  Class.\ Quant.\ Grav.\ }{{\bf #1} {(#2)} {#3}}}
\nc{\epl}[3]{{\it  Europhys.\ Lett.\ }{{\bf #1} {(#2)} {#3}}}
\nc{\ijmp}[3]{{\it Int.\ J.\ Mod.\ Phys.\ }{{\bf #1} {(#2)} {#3}}}
\nc{\ijtp}[3]{{\it Int.\ J.\ Theor.\ Phys.\ }{{\bf #1} {(#2)} {#3}}}
\nc{\jmp}[3]{{\it  J.\ Math.\ Phys.\ }{{ \bf #1} {(#2)} {#3}}}
\nc{\jpa}[3]{{\it  J.\ Phys.\ A\ }{{\bf #1} {(#2)} {#3}}}
\nc{\jpc}[3]{{\it  J.\ Phys.\ C\ }{{\bf #1} {(#2)} {#3}}}
\nc{\jap}[3]{{\it J.\ Appl.\ Phys.\ }{{\bf #1} {(#2)} {#3}}}
\nc{\jpsj}[3]{{\it J.\ Phys.\ Soc.\ Japan\ }{{\bf #1} {(#2)} {#3}}}
\nc{\lmp}[3]{{\it Lett.\ Math.\ Phys.\ }{{\bf #1} {(#2)} {#3}}}
\nc{\mpl}[3]{{\it  Mod.\ Phys.\ Lett.\ }{{\bf #1} {(#2)} {#3}}}
\nc{\ncim}[3]{{\it  Nuov.\ Cim.\ }{{\bf #1} {(#2)} {#3}}}
\nc{\np}[3]{{\it  Nucl.\ Phys.\ }{{\bf #1} {(#2)} {#3}}}
\nc{\pr}[3]{{\it Phys.\ Rev.\ }{{\bf #1} {(#2)} {#3}}}
\nc{\pra}[3]{{\it  Phys.\ Rev.\ A\ }{{\bf #1} {(#2)} {#3}}}
\nc{\prb}[3]{{\it  Phys.\ Rev.\ B\ }{{{\bf #1} {(#2)} {#3}}}}
\nc{\prc}[3]{{\it  Phys.\ Rev.\ C\ }{{\bf #1} {(#2)} {#3}}}
\nc{\prd}[3]{{\it  Phys.\ Rev.\ D\ }{{\bf #1} {(#2)} {#3}}}
\nc{\prl}[3]{{\it Phys\ Rev.\ Lett.\ }{{\bf #1} {(#2)} {#3}}}
\nc{\pl}[3]{{\it  Phys.\ Lett.\ }{{\bf #1} {(#2)} {#3}}}
\nc{\prep}[3]{{\it Phys\. Rep.\ }{{\bf #1} {(#2)} {#3}}}
\nc{\prsl}[3]{{\it Proc.\ R.\ Soc.\ London\ }{{\bf #1} {(#2)} {#3}}}
\nc{\ptp}[3]{{\it  Prog.\ Theor.\ Phys.\ }{{\bf #1} {(#2)} {#3}}}
\nc{\ptps}[3]{{\it  Prog\ Theor.\ Phys.\ suppl.\ }{{\bf #1} {(#2)} {#3}}}
\nc{\physa}[3]{{\it  Physica\ A\ }{{\bf #1} {(#2)} {#3}}}
\nc{\physb}[3]{{\it  Physica\ B\ }{{\bf #1} {(#2)} {#3}}}
\nc{\phys}[3]{{\it Physica\ }{{\bf #1} {(#2)} {#3}}}
\nc{\rmp}[3]{{\it  Rev.\ Mod.\ Phys.\ }{{\bf #1} {(#2)} {#3}}}
\nc{\rpp}[3]{{\it Rep.\ Prog.\ Phys.\ }{{\bf #1} {(#2)} {#3}}}
\nc{\sjnp}[3]{{\it Sov.\ J.\ Nucl.\ Phys.\ }{{\bf #1} {(#2)} {#3}}}
\nc{\spjetp}[3]{{\it Sov.\ Phys.\ JETP\ }{{\bf #1} {(#2)} {#3}}}
\nc{\yf}[3]{{\it Yad.\ Fiz.\ }{{\bf #1} {(#2)} {#3}}}
\nc{\zetp}[3]{{\it Zh.\ Eksp.\ Teor.\ Fiz.\  }{{\bf #1}  {(#2)} {#3}}}
\nc{\zp}[3]{{\it Z.\ Phys.\ }{{\bf #1} {(#2)} {#3}}}
\nc{\ibid}[3]{{\sl ibid.\ }{{\bf #1} {#2} {#3}}}

\nc{\rf}[1]{(\ref{#1})}
\nc{\nn}{\nonumber \\*}
\nc{\bfB}{\bf{B}}
\nc{\bfv}{\bf{v}}
\nc{\bfx}{\bf{x}}
\nc{\bfy}{\bf{y}}
\nc{\vx}{\vec{x}}
\nc{\vy}{\vec{y}}
\nc{\oB}{\overline{B}}
\nc{\oI}{\overline{I}}
\nc{\oR}{\overline{R}}
\nc{\rar}{\rightarrow}
\nc{\ti}{\times}
\nc{\slsh}{\hskip-5pt/}
\nc{\sm}{Standard~Model~}
\nc{\MP}{M_{\rm Pl}}
\nc{\tp}{t_{\rm Pl}}

\nc{\pmin}{p_{\rm min}}
\nc{\pmax}{p_{\rm max}}
\nc{\fo}{f_0}
\nc{\foi}{f_{0,i}\,}
\nc{\fop}{f_0^P}
\nc{\fou}{f_0^U}

\nc{\eff}{{\rm eff}}
\nc{\MT}{M_{\rm T}}
\nc{\ML}{M_{\rm L}}
\nc{\kk}{\vek{k}}
\nc{\pp}{{\rm p}}
\nc{\pt}{\partial_t}
\nc{\half}{{1\over 2}}
\nc{\w}{\omega}
\nc{\uhat}{\hat{U}_\w}

\nc{\etal}{\mbox{\it et al.}}
\nc{\ie}{{\it i.e. }}
\nc{\eg}{{\it e.g. }}
\nc{\trh}{T_{\rm RH}}

\begin{document}
\title{{\hfill {{\small  TURKU-FL-P37-01
        }}\vskip 1truecm}
{\LARGE Excited Q-Balls in the MSSM with gravity mediated supersymmetry breaking}
\vspace{-.2cm}}

\author{{\sc Tuomas Multam\"aki\footnote{email: tuomas@maxwell.tfy.utu.fi}}\\ 
\\{\sl\small Department of Physics, University of Turku, FIN-20014, FINLAND}}

\date{February 28, 2001}
\maketitle

\abstract{Excited Q-balls are studied by numerical simulations in the Minimal 
Supersymmetric Standard Model with supersymmetry broken by a gravity
mediated mechanism. It is found that there is a suppression factor of
$\cO(10^{-2})$ in the rate at which a Q-ball can emit their excess energy
compared to the rate set by the dynamical scale of the field, $m$. 
Furthermore, it is noted that a Q-ball can withstand a large amount of excess 
energy without
losing a significant amount of its charge. The cosmological importance of
these properties are considered for Q-balls in the thermal bath of the early universe.} 

\section{Introduction}
A scalar field theory with a global $U(1)$ symmetry can contain a
class of non-topological solitons, Q-balls \cite{cole2}.
A Q-ball is the ground state of the theory in the sector of fixed
charge so that its stability and existence are due to the
conservation of the $U(1)$ charge. In realistic theories, Q-balls 
are generally present in supersymmetric extensions of the Standard
Model \cite{kusenko405}. In particular the scalar potential 
of the Minimal Supersymmetric Standard Model (MSSM) has several flat 
directions \cite{dine} that can support lepton and/or baryon number
carrying Q-ball solutions in its spectrum.

In cosmology, Q-balls can play a significant role in the phenomenology
of the early universe. The Affleck-Dine condensate is unstable with
respect to fragmentation into Q-balls so that the small perturbations
naturally present in the AD-condensate lead to the formation of
Q-balls \cite{enqvist1,kusenko3}. The fate of the Q-balls then 
depends on the supersymmetry (SUSY)
breaking mechanism: if SUSY is broken by a gauge mediated mechanism they
can be absolutely stable and contribute to the dark matter content
of the universe \cite{kkst}, whereas if the SUSY breaking mechanism is gravity
mediated, the Q-balls will be short-lived but can still have 
interesting cosmological consequences, \eg they may explain the ratio 
of baryons to dark matter in the universe \cite{bbbdm}. 
The charge spectrum and the 
number of balls is dependent on the form of the potential and initial conditions of
the AD-condensate \cite{kawasaki1,kawasaki2,adnum}.

After the Q-balls have formed, they will be subjected to the thermal bath of
the early universe. The effect of the background clearly depends on how energetic 
the background radiation is and how strongly the Q-ball field couples to it.
The temperature of the universe is hence significant in determining the importance 
of the thermal background for Q-balls.

Thermal effects on Q-balls have been previously considered via two 
different mechanisms, dissociation \cite{enqvistdm,kusenko3} and
dissolution \cite{enqvistdm}. Dissociation of Q-balls proceeds through 
the collisions of thermal particles with the ``hard-core'' of a Q-ball
so that excess energy is transferred to the Q-ball. This excess
energy can then overcome the binding energy and dissociate the
Q-ball. Dissolution occurs due to the thermalization of the 
``soft edge'' of the Q-ball leading to charge transport from the Q-ball
to the background. In this work, dissolution and dissociation have been
considered in the gravity mediated scenario by studying the time
development of Q-balls with initial excess charge. The gauge mediated
case has been studied by thermodynamical means in \cite{laine}.

In addition to thermal erosion, Q-balls are also subjected to evaporation
\cite{cohen}, where Q-balls lose their charge via pair production at their
surface. The loss of charge through evaporation is generally much slower
than that due to thermal processes so evaporation becomes effective only after 
the universe has cooled enough for the thermal effects to be insignificant.

The scalar potential in the case of MSSM with SUSY broken by 
a gravity mediated scenario is given by (omitting the $A$-terms)\cite{enqvistdm}
\be{potEMcD} 
V(\Phi) = m^2|\Phi |^2(1 + K \log({|\Phi |^2\over M^2}))+
{\lambda^2\over M^{2(d-3)}_{Pl}}|\Phi|^{2(d-1)},
\ee
where $d$ is the dimension of the non-renormalizable term in the
superpotential, $m$ is the mass scale given by the SUSY breaking scale, $M$ 
is a large mass scale and $K$ a negative constant of order one.
The interesting cases for Q-balls are the d=4 $u^cu^cd^ce^c$
and d=6 $u^cd^cd^c$ directions in the superpotential.
Q-balls in this potential are well represented by a Gaussian
approximation \cite{enqvistdm}, where the Q-ball field is 
approximated by $\phi(t,\vec{x})=e^{i\w t}\phi_0 e^{-r^2/R^2}$. Here
$\w$ is the so called Q-ball frequency and $R\sim m^{-1}|K|^{-1/2}$. 
As can be seen from the form of the Gaussian approximation, Q-balls in the
gravity mediated case have thick walls. Their charge-energy
relation is well approximated by $E\approx mQ$.

\subsection{Dissociation and Dissolution}

A particle with energy $T$ can penetrate a Q-ball up to a distance
where its effective mass $g\phi(r)$ is equal to its energy \ie
$T\approx g\phi(r)$, where $\phi(r)$ is the Q-ball scalar field and 
$g$ the coupling of the thermal particle to the ball. The thermal particle
entering the ball will then lose some of its kinetic energy to the classical
Q-ball field and come to a temporary halt. The excess energy transferred
to the Q-ball can then be radiated as scalar field waves or lumps of
charge, or some combination of these. 

The rate of the number of particles hitting a Q-ball of radius $R$ in a thermal
path of temperature T is
\begin{equation}
{dN\over dt}={\tilde{g}(T)\over\pi^2}4\pi R^2T^3,
\end{equation}
where $\tilde{g}(T)$ is the effective number of thermal degrees of
freedom coupling to the Q-ball scalar. Typically we can set
$\tilde{g}(T)\approx 10^2$. 

The question of how much of the energy of an incoming particle
will be transferred to the Q-ball was considered in \cite{enqvistdm}. 
Assuming that the Q-ball field can reconfigure itself over a time scale
of $m^{-1}$, the Q-ball can typically absorb most of the energy
of the incoming particle. Hence, in each collision an energy of $\gamma_T T$,
where $\gamma_T\lsim 1$, is transferred to the Q-ball. The rate at which 
energy is being added to the Q-ball by the thermal background is therefore given by
\be{energyadd}
{dE\over dt}=\gamma_T T{dN\over dt}={\tilde{g}(T)\over\pi}4\gamma_T R^2T^4.
\ee

To avoid dissociation, a Q-ball should be able to radiate the excess
energy as scalar field waves at a rate faster than given by
(\ref{energyadd}). Otherwise, energy will be accumulated in the Q-ball 
until it begins to dissociate.

In addition to dissociation, a Q-ball can lose its charge by dissolution
where charge is removed from the ``soft edge'' of the Q-ball. 
For $T\gsim m$, the thermal
background quarks have a much smaller mean free path,
$\lambda_{mfp}\approx k_q/T\ (k_q\approx 6)$  than 
the width of the soft edge of a Gaussian Q-ball, 
$\delta r\approx R/(2\beta)$, where  
$\beta=\ln(g\phi_0/3T)^{1/2}$.
The charge contained in the soft edge is approximately
$Q_{\rm soft}\approx 4\pi\w (T/g)^2(\beta R)^2\delta r$
\cite{enqvistdm}. The background quarks
can therefore thermalize the soft edge of a Q-ball at high temperatures.
The time taken for a charge to leave the soft edge by a random walk process
is \cite{enqvistdm}
\be{taud}
\tau_d\approx ({R\over 2\beta})^2{T\over k_q}\approx {T\over 4\beta^2 m^2|K|k_q}.
\ee
The rate at which charge can be removed then depends on how quickly the
Q-ball can reconfigure itself and replenish charge in the soft edge. If
the charge is replenished quickly enough, the rate
of charge loss is $dQ/dt\approx Q_{soft}/\tau_d$, where $Q_{soft}$
is the charge contained in the soft edge. However, if the Q-ball
does not reconfigure and hence replenish the charge quickly enough,
the rate of charge loss will be suppressed.

\section{Numerical simulations}

\subsection{Preliminaries}

To study the dynamics of excited Q-balls, they have been 
simulated numerically on a 2+1 dimensional lattice. 
Obviously, a 3+1 dimensional simulation would be more realistic 
but it is expected that a 2+1 dimensional simulation gives a fair approximation
to the more realistic case while greatly reducing the required CPU time.

The equation of motion in flat space corresponding to the
potential (\ref{potEMcD}) is 
\be{eqm} 
\ddot\Phi+\nabla^2\Phi+m^2\Phi[1+{K\over \log(10)}+K\log({|\Phi|^2\over M^2})]+
{2(d-1)\lambda^2\over\MP^{2(d-3)}}|\Phi|^{2(d-2)}\Phi=0.
\ee 
(We can neglect the expansion of the universe for our purposes since we
are interested in the behaviour of a single, isolated Q-ball.)

Since our goal is to address questions relevant to Q-balls subjected
to a thermal background radiation, it is necessary to consider exactly 
what to simulate while taking into account the numerical and practical 
problems that may arise. 

Simulating the collision of a single particle of energy $T$
with the Q-ball is cumbersome due to the vastly different energy scales
involved. In a realistic cosmological setting, the energy of a Q-ball is 
typically much larger than $T$. 
Keeping track of such a tiny change in energy would
require a simulation where the Q-ball field is let to evolve
at very high accuracy while keeping the time and spatial steps 
very small. This in order requires a large lattice and several orders
of magnitude more time steps. Also simulating a single particle
hitting the Q-ball is not realistic since a Q-ball in the thermal
bath of the early universe is subjected to several particles 
interacting with it simultaneously.

Another approach would be to fully simulate both a generic background field and the
Q-ball field. The difficulty in this approach is to keep the background
field at a constant temperature at large distances \ie to keep the
system from reaching a global equilibrium state due to the smallness
of the lattice compared to the Q-ball size. By using a much larger
lattice this approach would probably be more feasible but would
again require more CPU time. Hence an approximate method has been
adopted: instead of simulating the whole thermal erosion process,
we simply study the evolution of a Q-ball with excess energy added
in the form of a random perturbation, $\phi=\phi_0+\delta\phi$,
where $\phi_0$ is the Q-ball field and $\delta\phi$ is a stochastic term.
Such a perturbation adds to the energy of the Q-ball while keeping the contribution to the
charge small. The exact form of the perturbation is not important;
the relaxation times are insensitive of how the extra energy is added to the
Q-ball. For example, if energy is added asymmetrically, it has been
noted that the fluctuations of the field quickly distribute around the 
Q-ball before excess energy begins to radiate.

After the perturbation is added to the initial configuration,
the field is allowed to develop until the Q-ball becomes stable again
\ie excess energy and possibly charge is radiated from the ball.
To avoid any excess energy and possible charge radiated by the Q-ball 
from entering the system due to continuous boundary conditions, the lattice
has dissipative zones near the boundaries (to minimize back-scattering
from the boundaries of the dissipative regions the dissipation term
is increased gradually). For the numerical simulations 
the field was decomposed into real and
imaginary parts, $\Phi={1\over\sqrt{2}}(\phi_1+i\phi_2)$, and re-scaled
along with the spatial and temporal unit, 
$\varphi_i={\phi_i\over\sqrt{2}M},\ \tau=mt,\ \xi=mx$. The parameter values we used
in the numerical simulations were $m=100\GeV,\ K=-0.1,\ M=10^9\GeV$.
The calculations were done mainly for the $d=4$ case but we have checked that 
the $d=6$ case does not significantly differ from these results.
We have also verified that varying $K$ in the typical range $-(0.01-0.1)$
does not markedly affect the relaxation rate. The lattice size was 
typically $200\times 200$ and the lattice parameters 
were chosen as $\Delta\xi=0.6,\ \Delta\tau=0.3$ (a number of 
simulations with varying values of $\Delta\xi$ and $\Delta\tau$ were
done to ensure that the exact values of the lattice parameters
do not affect the results).

\subsection{Numerical results}

We have studied numerically the relaxation of Q-balls for 
different Q-ball charges and excess energies.
For each set of parameter values a number of simulations were run
to account for the effect induced by the stochastic term \ie
a different seed for the pseudo random number generator in the
stochastic term was used. 

\begin{figure}
\leavevmode
\centering
\vspace*{90mm}
\includegraphics{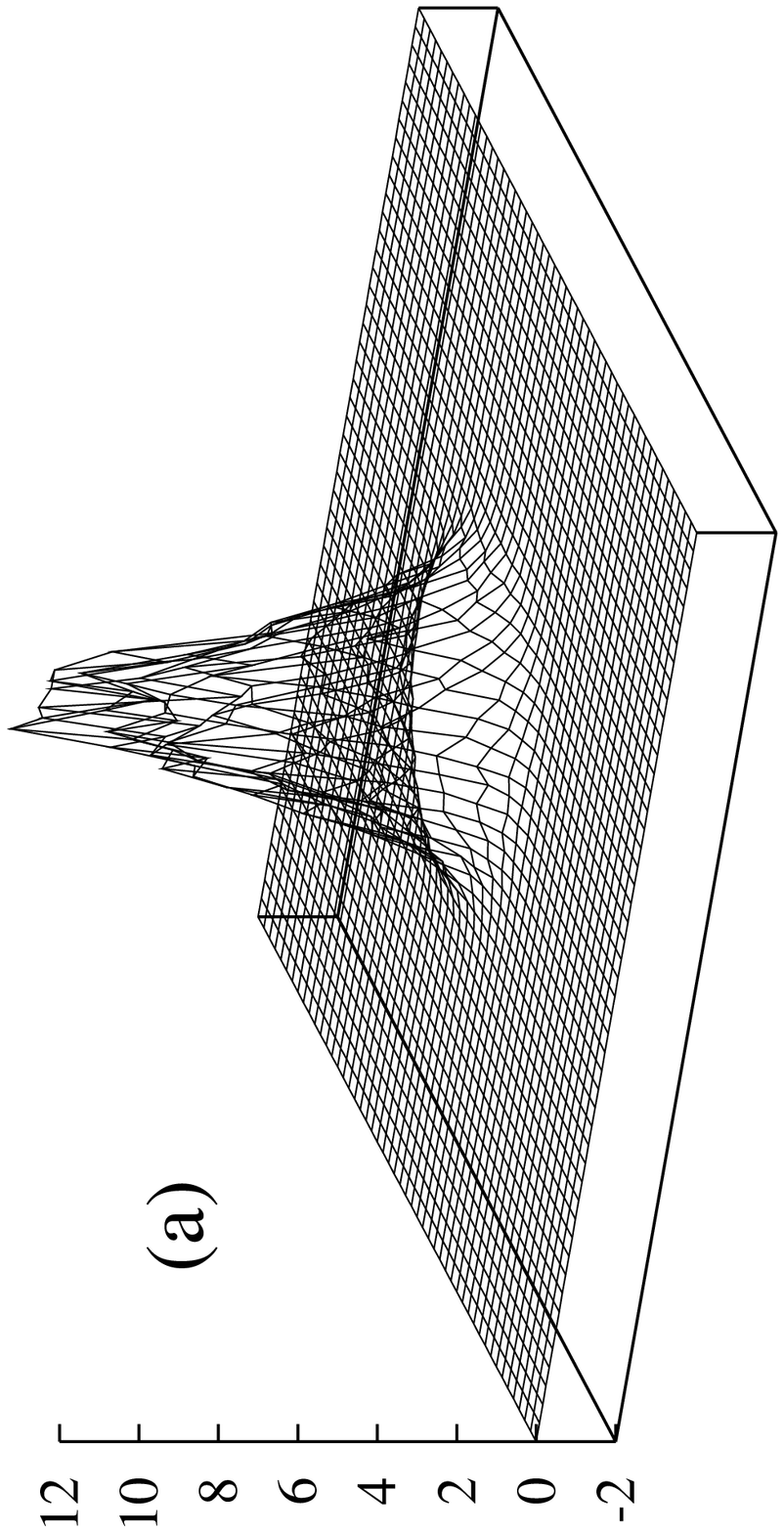}
\includegraphics{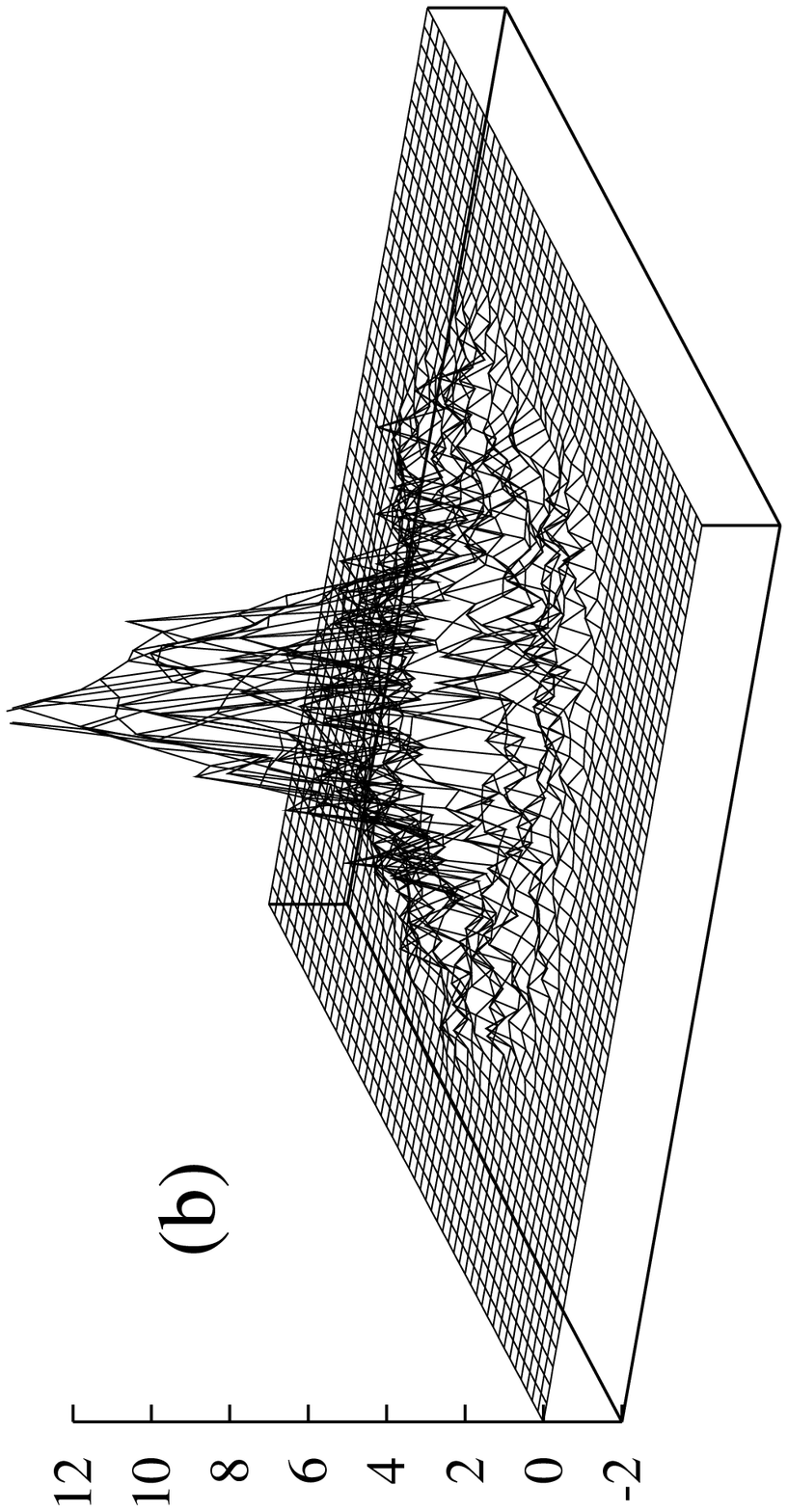}
\includegraphics{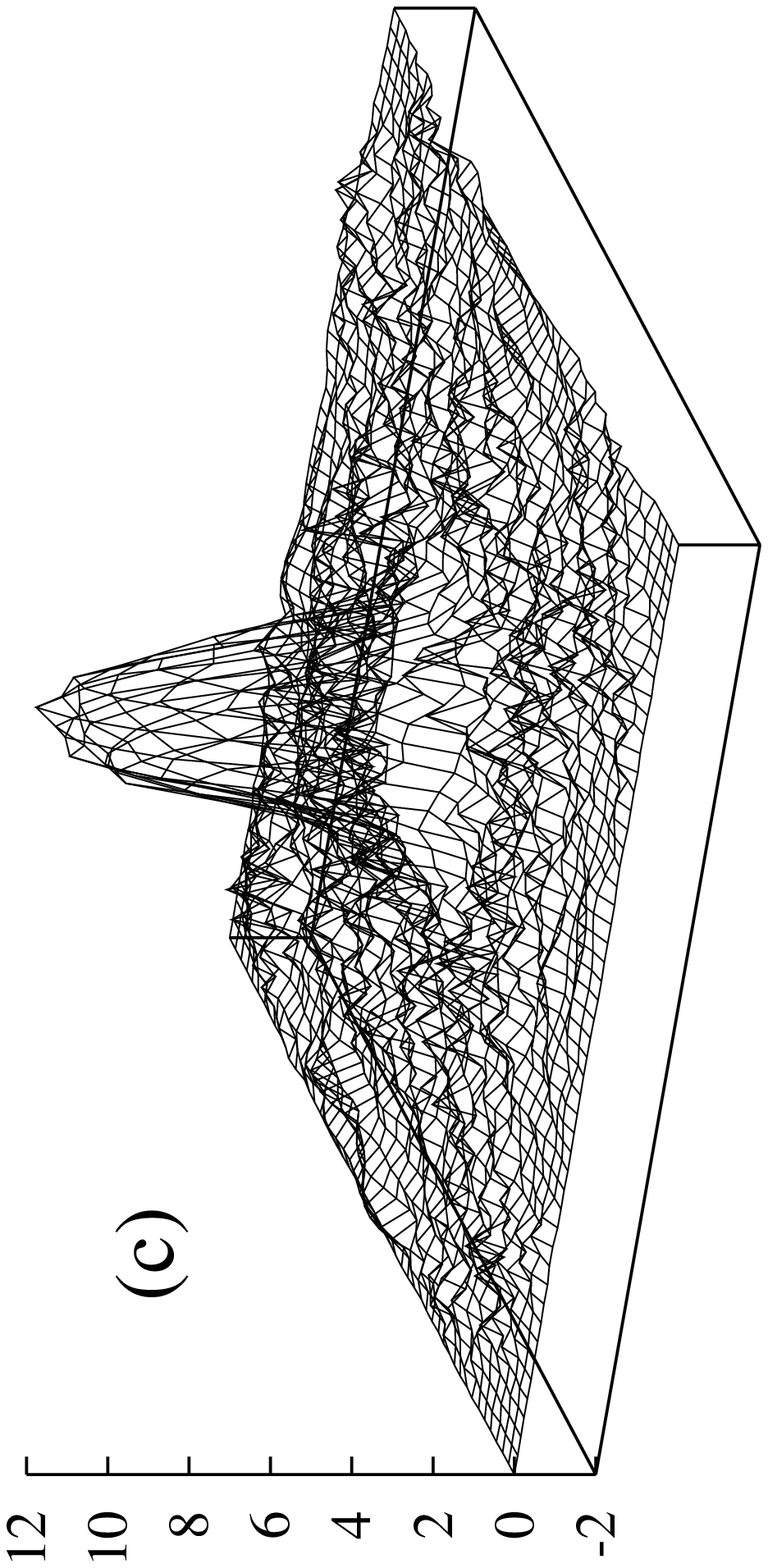}
\includegraphics{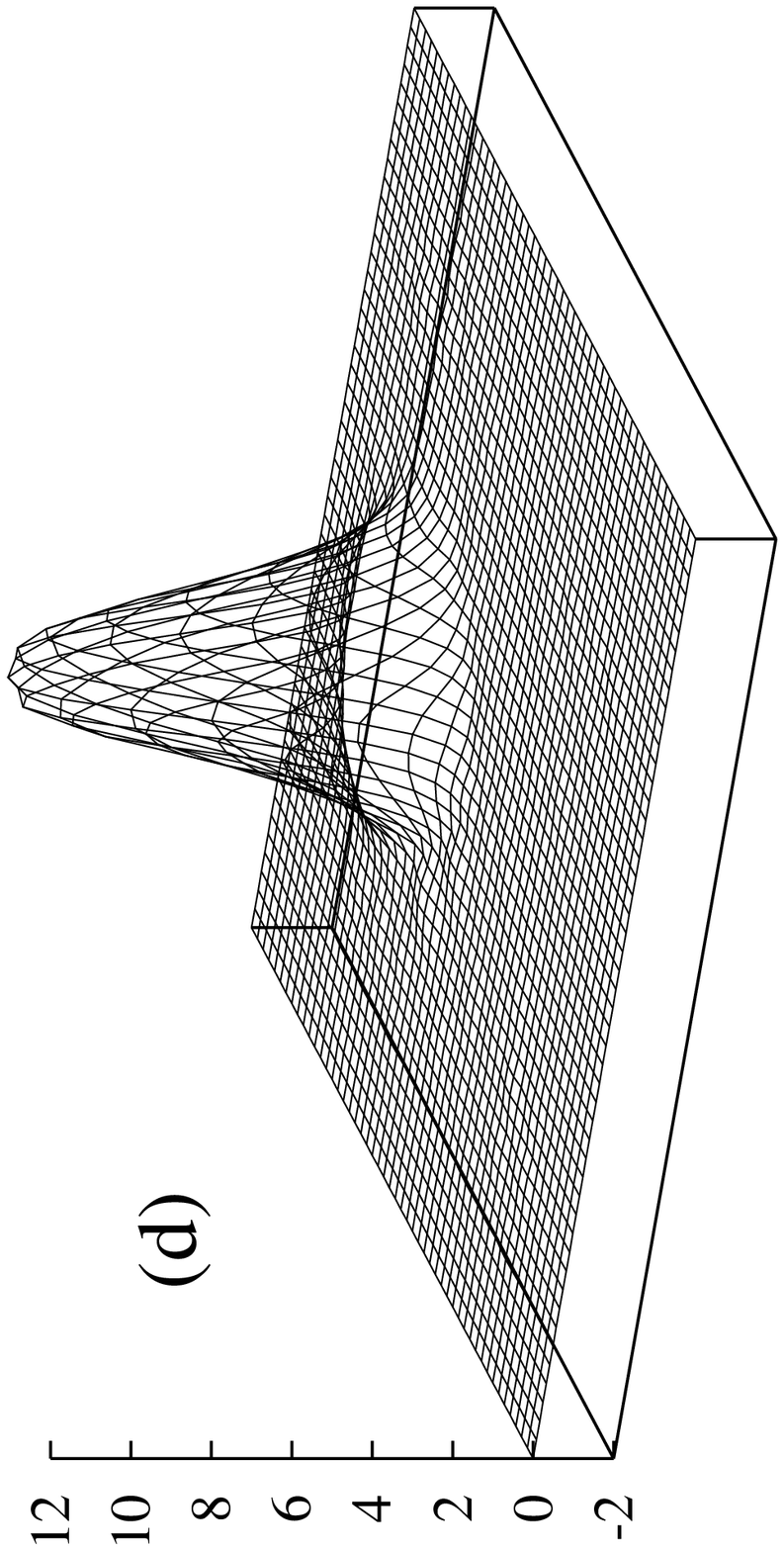}
\caption{A typical Q-ball relaxation process}\label{typical}
\end{figure}

A typical relaxation process is shown in Fig. \ref{typical} where
the amplitude of the scalar field is plotted. 
The initial configuration, with $\w=0.80m$ corresponding to 
$Q=9100$ and $E=7600$ 
(in lattice units), is shown in Fig. \ref{typical}(a),
is put on a $200\times 200$ lattice and the field is let
to evolve. Note that the images are not taken at equal time intervals:
the time units at which snapshots are taken are
$\tau_a=0,\ \tau_b=20,\ \tau_c=40,\ \tau_d=990$.
From the figures it can be seen how the Q-ball quickly radiates the excess energy 
in the form of scalar field waves which propagate spherically from the
ball. The waves are dissipated at the dissipation zones as they reach 
the boundaries of the lattice. Radiation gradually decreases
and in the last panel one can see virtually no radiation. From Fig.
\ref{typical}(d) one can also note how the Q-ball has slightly moved
from its initial position in the center of the lattice due to a
spatial asymmetry present in the initial conditions.

\begin{figure}
\leavevmode
\centering
\vspace*{60mm}
\includegraphics{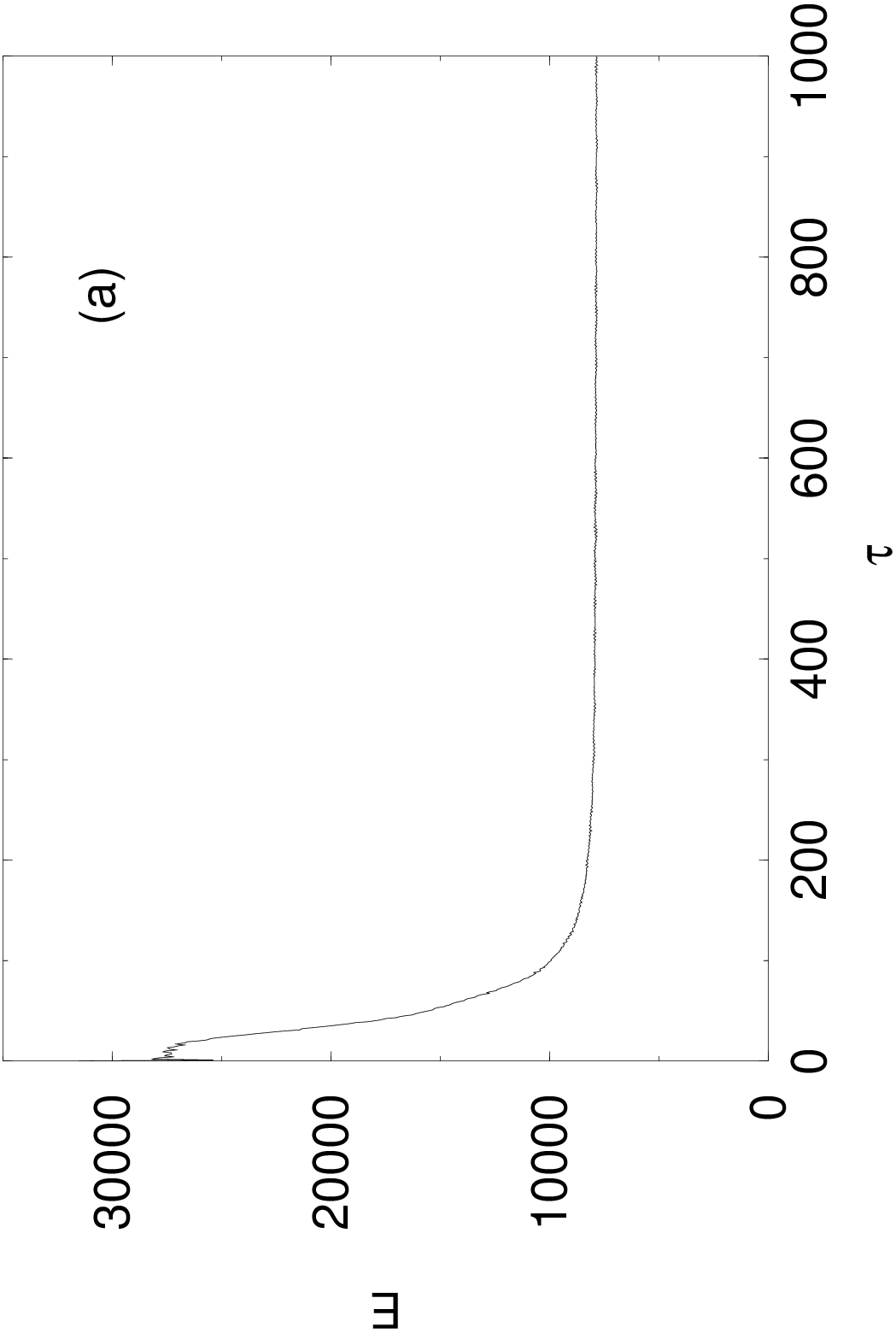}
\includegraphics{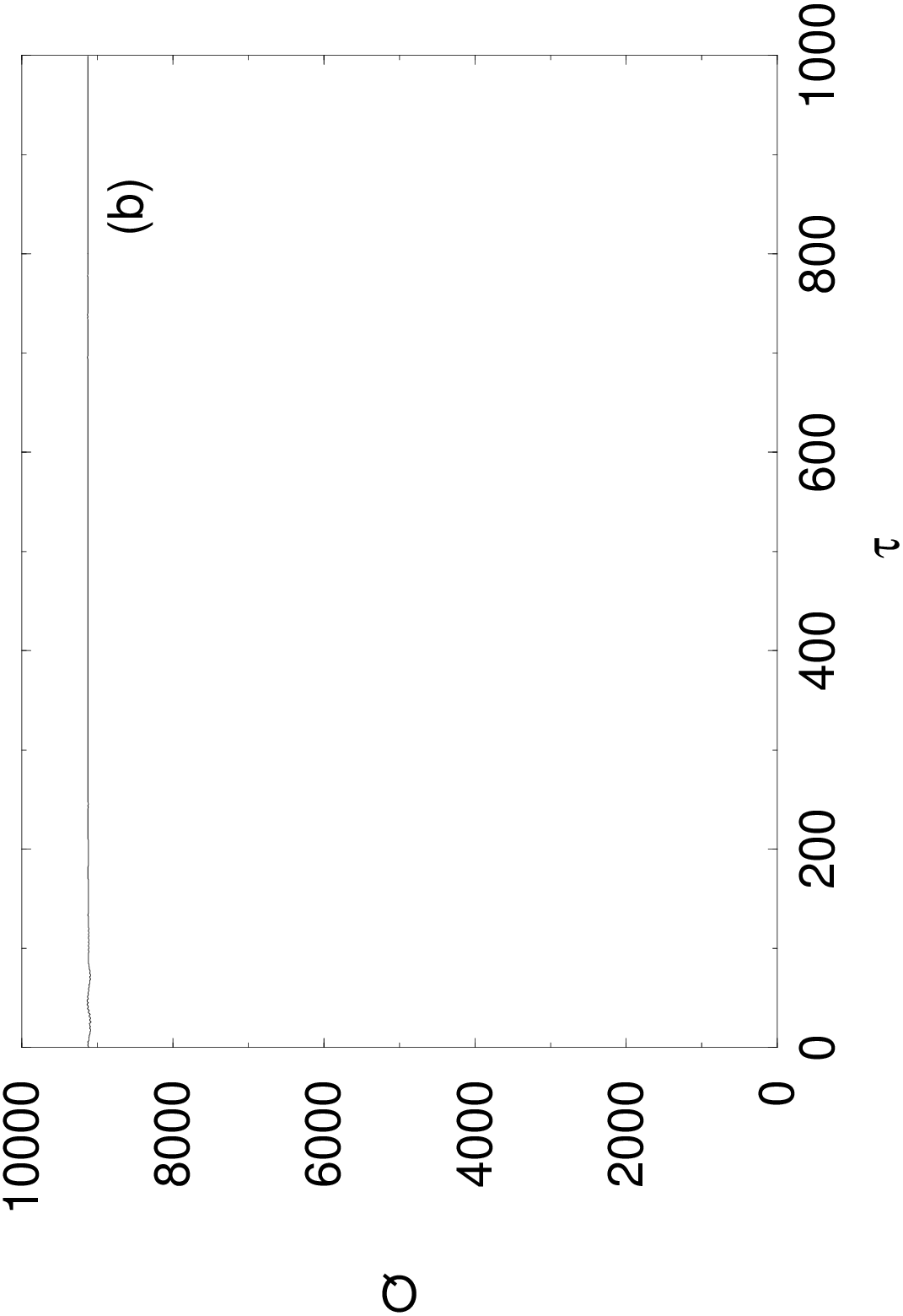}
\includegraphics{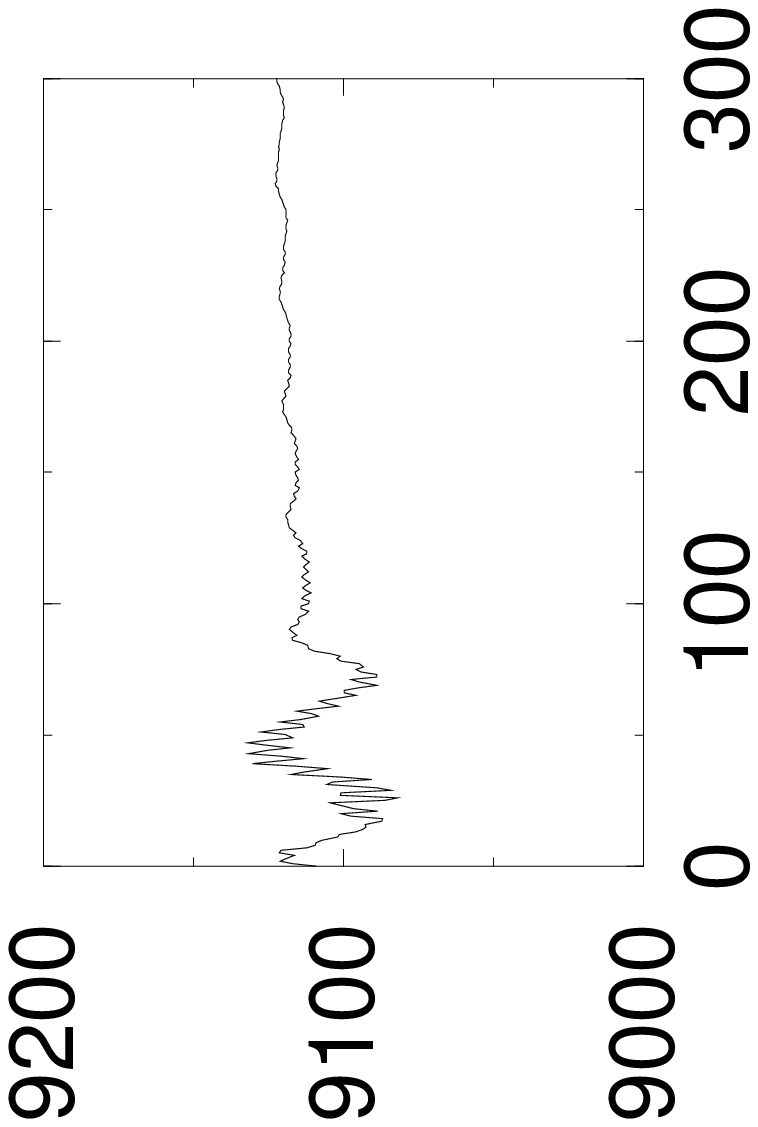}
\caption{Q-ball energy (a) and charge (b) as a function of $\tau$, $\w=0.80m$}\label{eandq}
\end{figure}

The energy and charge of the Q-ball in lattice units have been plotted in
Fig. \ref{eandq}. From the figure it can be seen how energy
is emitted from the ball while the charge remains virtually
constant, even though a significant amount of excess charge is
introduced to the Q-ball initially. The slight fluctuation of charge at the
beginning of the simulation is shown in the close up in Fig. \ref{eandq}(b).

To study how the energy is emitted from the Q-ball
a number of different functions has been fitted to the energy curves.
It is found that the energy curve is well described by
\be{efit}
E(\tau)=A[\theta(\tau_0-\tau)+\theta(\tau-\tau_0)\e^{-\lambda(\tau-\tau_0)}]+E_0,
\ee
where $A,\lambda, E_0$ are the fitting parameters and 
$\theta(\tau)$ is the step function. The presence of a non-zero
$\tau_0$-term implies that the excess energy first spreads around 
the ball before it is emitted.

\begin{figure}
\leavevmode
\centering
\vspace*{60mm}
\includegraphics{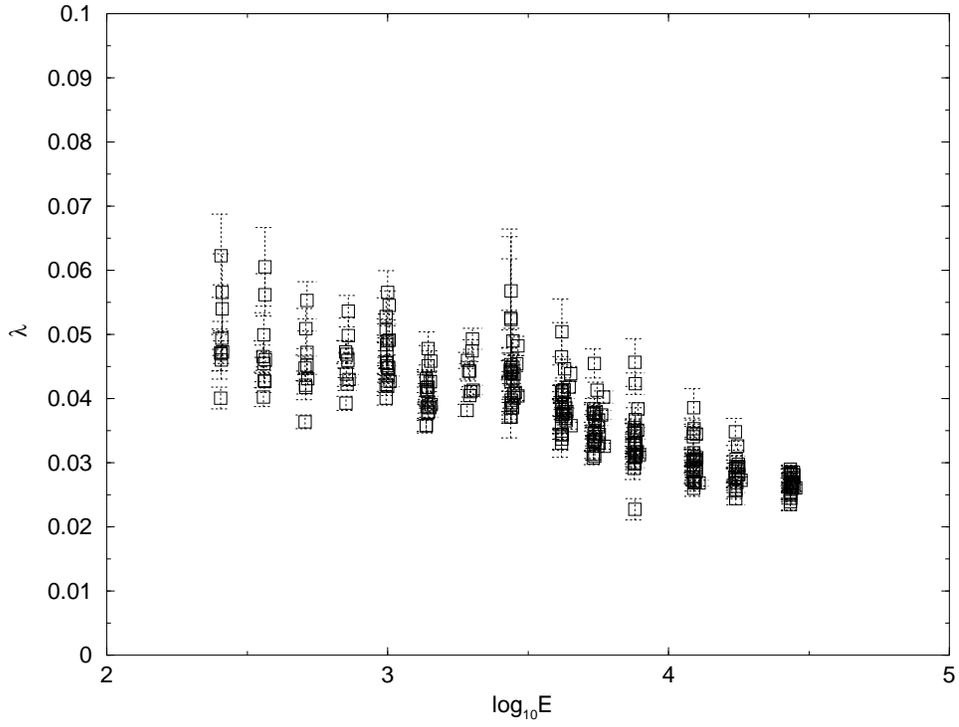}
%\special{psfile=betazoom.ps voffset=70 hoffset=0 vscale=20
%hscale=20 angle=-90}
\caption{$\lambda$ for $\w=0.78m-0.98m$ as a function of Q-ball energy}\label{betas}
\end{figure}

The rate at which excess energy is emitted has been studied for $\w=0.78m-0.98m$,
which in two spatial dimensions corresponds to Q-ball charges in the range
$Q\sim 10^{16}\ -\ 10^{18}$ (note that smaller values of $\w$ correspond to larger charges).
The values of $\lambda$ have been plotted in Fig. \ref{betas} as a
function of the unperturbed Q-ball energy (in lattice units). 
For each value of $\w$ the maximum amplitude of the perturbation $\delta\phi$
was varied, $|\delta\phi|\leq a\phi(0)$, where $a$ was varied in the range
$0.01-0.10$. The maximum value of $a$ for each $\w$ was chosen 
so that the charge of the Q-ball varied less than $1\%$ from its initial
value during the relaxation process. Ten simulations were run with a 
different initial stochastic perturbation for each $a$.

The amount of excess charge is obviously a function of $a$ and hence 
it varies greatly for each value of $\w$.
For example, for the $\w=0.78m$ case, the values of $\lambda$ have 
been calculated for increases in energy between $1\%$ and $40\%$ of 
the energy of the ground state. 
However, as from Fig. \ref{betas} can be seen, the variations in $\lambda$ 
are small compared to the changes in energy. The values of $\lambda$ 
increase slightly with smaller energy so that smaller Q-balls radiate their 
excess energy more quickly than larger Q-balls.

\begin{figure}
\leavevmode
\centering
\vspace*{0mm}
\includegraphics{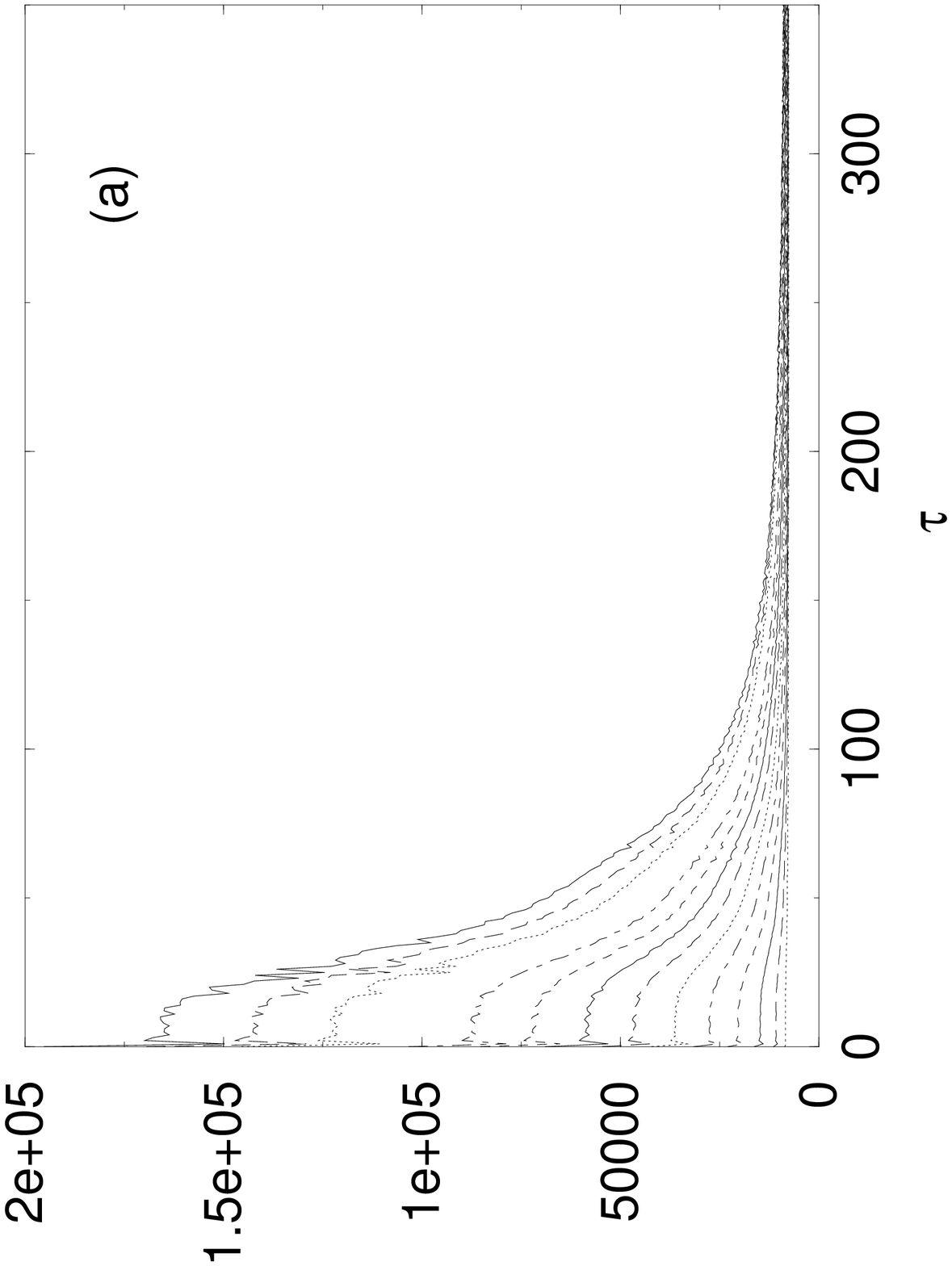}
\includegraphics{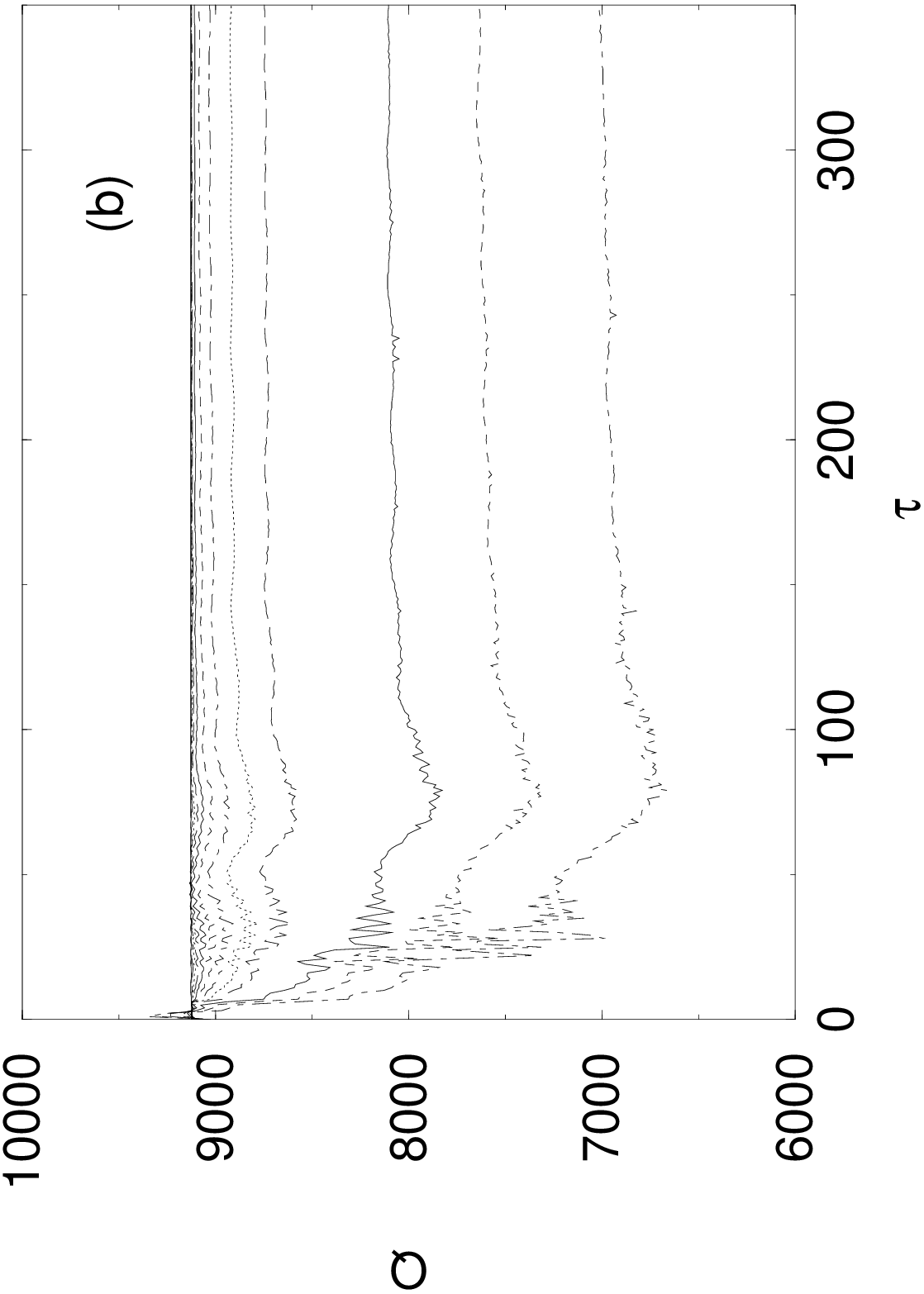}
\caption{Q-ball energy (a) and charge (b), $\w=0.80m$, $Q_0=9100$, $E_0=7600$}
\label{makseq}
\end{figure}

We have also studied the effect of adding larger amounts of excess
energy to a Q-ball. The time development of energy and charge of a 
Q-ball with perturbations of different magnitude for $\w=0.80m$ is 
shown in Fig. \ref{makseq}. The ground state has charge $Q_0=9100$ and
energy $E_0=7600$. From the figure it can be seen how energy still 
decreases exponentially even with very large initial excess energy. 
As the amount of excess energy grows, the Q-ball emits more charge 
along with the energy but still retains most of its charge in a single Q-ball.
In the case shown in Fig. \ref{makseq}, more than $75\%$ of the
initial charge remains in the Q-ball even if the excess 
energy is $\sim 10^5$. A Q-ball can hence survive fragmentation 
even with initial excess energies of the order of $\sim 10$ times 
its initial energy.

\section{Discussion}

From the numerical results we have seen that the energy loss
experienced by an excited Q-ball is well described by an exponential
function, $\dot{E}_Q(t)=-\lambda(E_Q(t)-E_{Q_0})$, 
where $E_{Q_0}$ is the ground state of the Q-ball.

To study the dissociation of Q-balls in the early universe we need
to consider the energy increase due to the thermal background.
Since the rate at which energy is being added to the
Q-ball due to the thermal background is given by (\ref{energyadd}), 
we can model the energy of the Q-ball in the early universe by
\be{qedy}
\dot{E}_Q(t)=-\lambda(E_Q(t)-E_{Q_0})+{\tilde{g}(T)\over\pi}4\gamma_TR^2T^4.
\ee
Here we can approximate that $R$ is approximately independent of the 
charge (and energy) of the Q-ball, assuming that the energy increase
is not large enough to dissociate the Q-ball.

The temperature, $T$, decreases with time due to the expansion the 
universe so that in a radiation dominated universe $T^2t\sim\MP$ 
and in a matter dominated universe $T^{3/2}t\sim\MP$.
In general one can therefore write (\ref{qedy}) as
\be{qedy2}
\dot{E}_Q(t)=-\lambda(E_Q(t)-E_{Q_0})+{A\over t^B},
\ee
where $A$ is a constant and $B=2$ (radiation dominated) or $B={8\over 3}$
(matter dominated). Integrating once with respect to time one gets
\be{qedyint}
E_Q(t)=E_{Q_0}+(E_Q(t_0)-E_{Q_0})\e^{-\lambda(t-t_0)}+A\e^{-\lambda t}
\integral{t_0}{t}dx\ {\e^{\lambda x}\over x^B},
\ee
where $t_0$ is now the time at which the Q-ball is formed.

The relevant features of the time evolution of Q-ball energy can be
extracted from (\ref{qedy}). As $t\rar\infty$, the energy
of a Q-ball tends to the ground state (assuming that the Q-ball does not
dissociate during the process) \ie $E_Q(t\rar\infty)=E_{Q_0}$.
If there exists an extremum $t_c$, $\dot{E}(t_c)=0$, any subsequent
extremum points, $\dot{E}(t)=0$, will occur at a lower energy \ie
$E(t_c)>E(t)$ for $B>1$. Hence the value of $E(t_0)$ determines
the features of the time evolution of a Q-ball: If 
$E_Q(t_0)<A/(\lambda t_0^B)+E_{Q_0}$, $\dot{E}(t_0)$ is positive and the 
Q-ball energy will initially increase until the energy loss from the Q-ball
exceeds the contribution from the background. The increase in energy is 
clearly bounded from above by $At_0^{1-B}/(B-1)$. If, on the other hand,
$\dot{E}(t_0)<0$, the energy of a Q-ball will monotonously decrease with time.

If one assumes a situation where Q-balls are formed before 
reheating due to the inflaton decays and are approximately at their
ground state, $E_Q(t=t_0)\approx E_{Q_0}$, the energy of a Q-ball 
will initially increase, $\dot{E}(t)>0$. The temperature of the universe after 
reheating can be approximated by $T(t)=\sqrt{\MP\over t},\ t\geq t_{\rm RH}=
\MP/\trh^2$, where $\trh$ is the reheating temperature and $t_{RH}$ is 
the reheating time. The maximum energy of a Q-ball clearly depends on 
the exact value of parameters. However, for typical parameter values, 
numerical work shows 
that for $\trh\sim 10^2-10^9\GeV$ 
the increase in energy is well approximated by 
\be{emax}
E_Q^{\rm max}(t>t_{\rm RH})\approx E_Q(t_{\rm RH})+{\tilde{g}(T)\over\pi\lambda}4\gamma_TR^2T_{RH}^4,
\ee
which corresponds to the point at which the rates at which energy is 
emitted and absorbed are equal. Clearly, (\ref{emax}) is 
valid only if the Q-balls emit their excess charge quickly enough
compared to the rate of energy increase due to the energetic background.

From the simulations we know that to significantly remove charge
from a Q-ball we need to add a perturbation whose energy is
comparable to the energy of the Q-ball itself. Therefore, to avoid 
significant loss of charge the energy of a Q-ball is bounded from below by
\be{qlimit}
E_Q\approx mQ\gsim {\tilde{g}(T)\over\pi\lambda\alpha}4\gamma_TR^2T_{RH}^4,
\ee
where $\alpha$ is the fraction of excess energy that is required to 
significantly remove charge from the Q-ball.
Using $R\sim m^{-1}|K|^{-1/2}$, $K=-0.1$, $\gamma_T\approx 1$ (\ie all
of the energy of an incoming particle is transferred to the Q-ball), $\tilde{g}(T)\approx 100$,
$\lambda\sim 0.02m$ and $\alpha\sim 1$ we get
\be{nqlimit}
T_{RH}\lsim 6\times Q^{1/4}\GeV.
\ee

On the other hand, if Q-balls are still at an excited state 
at the time of reheating and the reheating 
temperature is low, the rate of energy loss from a Q-ball can be greater than 
the rate of energy increase due to reheating. The energy 
of a Q-ball can then be monotonously decreasing with time during the
reheating process and dissociation due to the energetic background 
will not play a major role in the evolution of Q-balls. 
 
In addition to dissociation, dissolution can also be a significant factor
in determining the fate of Q-balls in the thermal bath of the early
universe. For dissolution to be important the rate of charge loss
from the soft edge, ${\tau_d}^{-1}$, must be smaller than the rate
at which charge is replenished. The rate which a Q-ball can reconfigure
itself and replenish charge in the soft edge now becomes important.
The numerical simulations show that the Q-ball emits excess charge
at a slow rate, $\Gamma\sim\lambda m$ compared to dynamical scale of the scalar field,
$m$. One can then speculate that there is a suppression factor
in how quickly charge is replenished in the soft edge, $\tau_r\sim (\kappa m)^{-1}$.

For efficient dissolution we need
$\tau_d^{-1}\lsim\tau_r^{-1}$ so that using (\ref{taud}) we get a
lower bound for the background temperature for ,
\be{dissolcond} 
T\gsim 4\beta^2m|K|k_q \kappa^{-1}.
\ee
Using typical values, we get $T\gsim {4\over\kappa}\times 10^3\GeV$
so that the presence of a suppression factor can significantly
increase this bound. For example, assuming that $\kappa\sim\lambda$, 
we get $T\gsim 10^5\GeV$.

At temperatures high enough for efficient dissolution, charge in 
the soft edge is replenished quickly enough and dissolution is efficient.
The rate of charge loss is then given by,
\be{dissolloss}
{dQ\over dt}\approx {Q_{\rm soft}\over \tau_d}=8\pi\w\beta^3k_qRg^{-2}T,
\ee
where (\ref{taud}) has been used.
Using $T=(\MP/2k_Tt)^{1/2}$ and integrating, the total loss of charge
from the time of reheating until temperature has decreased to $T$,
is given by
\be{totloss}
\Delta Q=8\pi\w\beta^3k_qRg^{-2}k_T^{-1}\MP({1\over T}-{1\over T_{RH}}).
\ee
The temperature at which the loss of charge due to dissolution 
is no longer efficient is given by (\ref{dissolcond}) and hence
(\ref{totloss}) is bounded from above by
\be{maxdeltaq}
\Delta Q\leq {2\pi\w\beta\kappa\MP\over g^2 k_T m^2 |K|^{3/2}}.
\ee
For typical values $\w\sim m,\ \beta\sim 4,\ g\sim 1, k_T\approx 17$ 
and $K=-0.1$ we get $\Delta Q\sim 0.5\kappa(\MP/{\rm GeV})$.
One should also note that the rate at which charge leaves the soft edge
of a Q-ball by random walk will in reality be less than ${\tau_d}^{-1}$
since random walk transports charges in all spatial directions and 
the rate at which charge drifts away from the Q-ball will depend
on the gradient of the charge density. Furthermore, charge diffusion 
away from the edge of the Q-ball needs to be efficient to prevent 
charges being transported back into the Q-ball. These considerations
can further suppress the dissolution process.

\section{Conclusions}

We have studied the effect of adding excess energy to a Q-ball by
numerical means. From doing a set of simulations we have found
that the excess energy is lost as $E(t)\sim e^{-\lambda\tau}\sim
e^{-\lambda m t}$, where $\lambda\sim 0.02-0.06$ in the studied parameter range. 
From naive arguments one would expect that the relaxation rate of the
Q-ball would be of the order of $m$. The numerical simulations
show that there is a suppression factor $\cO(10^{-2})$
in how quickly Q-balls radiate their excess energy. We expect that
this suppression factor would also be present in full 3D simulations
even though the Q-balls are likely to radiate their excess energy
slightly more quickly due to the presence of an additional degree 
of freedom. We have also observed the slow relaxation rate of excited Q-balls 
when studying Q-ball collisions \cite{gravcolli}. 

In addition to the fact that Q-balls tend to radiate their excess energy
more slowly than what one might expect, it has also been found that 
the Q-balls are surprisingly robust. Even when the excess
energy added is a significant fraction of the energy of the unperturbed
Q-ball and clearly exceeds the binding energy, $mQ-E$, only a very
small fraction of the charge is lost in emitting the excess energy.
Only when the excess energy is much larger than the Q-ball energy
we start to see a significant loss of charge. One can gain an insight 
into this by looking at the effective potential of a single ``free'' 
$\phi$ particle inside the Q-ball: the particle is typically in a deep 
potential well with a high barrier that also suppresses quantum tunneling.

On the basis of the small rate of radiation of the excess energy,
one can speculate that also the reconfiguration rate may be suppressed.
This would then reduce the rate at which a Q-ball can absorb
energy from the thermal background and suppress the charge loss
by dissolution. Especially the effect on dissolution is significant
since a suppression factor in the reconfiguration rate
affects linearly the lower bound for efficient dissolution (\ref{dissolcond})
and hence also the upper bound for charge loss (\ref{maxdeltaq}).

The slow relaxation rate and the robustness of Q-balls have an
effect on how well they can avoid thermal erosion in the
early universe. The slow relaxation rate allows for the excess 
energy to build up in a Q-ball but on the other hand the excess 
energy required to significantly reduce the charge of Q-ball 
by dissociation is large. In a realistic cosmological setting one 
should also take into account the effect of preheating due to inflaton 
decays and the decays of the remains of the AD-condensate. However, on 
the basis of the arguments presented in \cite{enqvistdm}, preheating 
effects should not significantly affect the bounds for the Q-ball charge 
and reheat temperature.

Q-balls formed from the Affleck-Dine condensate can survive the
thermal background of the early universe if the reheat temperature
is not too large. On the basis of the results presented in this work
Q-balls with charges $\sim 10^{20}$ should survive reheat temperatures
around $T_{\rm RH}\sim 10^5 \GeV$. The question of how quickly a Q-ball 
can absorb energy and replenish charge in its soft edge can, however,
change theses limits so further work is still needed to understand
the exact dynamics of how Q-balls erode thermally.

%%%%%%%%%%%%%%%%%%%%%%%%%%%%%%%%%%%%%%%%%%%%%%%%%%%%%%%%%%%%%%%%%

\section*{Acknowledgments}
I. Vilja is thanked for helpful discussions and careful reading
of the manuscript. This work has been supported by the Finnish
Graduate School in Nuclear and Particle Physics.

%%%%%%%%%%%%%%%%%%%%%%%%%%%%%%%%%%%%%%%%%%%%%%%%%%%%%%%%%%

\newpage

\newpage

\end{document}